\documentclass[11pt, a4paper]{article}
\pdfoutput=1
\usepackage[margin=2cm]{geometry}
\usepackage{amsmath, amsfonts, amssymb}
\usepackage{multirow}
\usepackage{graphics}
\usepackage{pgf}
\usepackage{verbatim}
\usepackage[utf8]{inputenc}
\usepackage[colorlinks=true
  ,urlcolor=blue
  ,anchorcolor=blue
  ,citecolor=blue
  ,filecolor=blue
  ,linkcolor=blue
  ,menucolor=blue
  ,pagecolor=blue
  ,linktocpage=true
  ,pdfproducer=medialab
  ,pdfa=true
]{hyperref}

\usepackage{cite}

\newcommand{\RDp}{\ensuremath{ R_{D^{(*)}}} }
\newcommand{\RKp}{\ensuremath{ R_{K^{(*)}}} }

\newcommand{\Rks}{\ensuremath{R_{K^{*0}}}}
\newcommand{\GeV}{\ensuremath{\,{\rm GeV}}}
\newcommand{\Clq}{\ensuremath{ C_{\ell q} } }
\newcommand{\Clqt}{\ensuremath{ C_{\ell q(3)} } }
\newcommand{\Clqo}{\ensuremath{ C_{\ell q(1)} } }
\newcommand{\CneNP}{\ensuremath{C_9^{\rm{NP}\, e}}}
\newcommand{\CteNP}{\ensuremath{C_{10}^{\rm{NP}\, e}}}
\newcommand{\CnmuNP}{\ensuremath{C_9^{\rm{NP}\, \mu}}}
\newcommand{\CtmuNP}{\ensuremath{C_{10}^{\rm{NP}\, \mu}}}

\begin{document}
\thispagestyle{empty}

\vspace*{2.5cm}

\vspace{0.5cm}

\begin{center}

\begin{Large}
\textbf{\textsc{Anomalies in $B$ mesons decays:\\ Present status and future collider prospects.}}
\end{Large}

\vspace{1cm}

{\sc
{\underline{J. Alda}}$^{a, b}$%
\footnote{\tt \href{mailto:jalda@unizar.es}{jalda@unizar.es}}%
, J.~Guasch$^{c}$%
\footnote{\tt \href{mailto:jaume.guasch@ub.edu}{jaume.guasch@ub.edu}}%
, S.~Pe{\~n}aranda$^{a, b}$%
\footnote{\tt \href{mailto:siannah@unizar.es}{siannah@unizar.es}}%
}

\vspace*{.7cm}

{\sl
$^a$Departamento de F{\'\i}sica Te{\'o}rica, Facultad de Ciencias,\\
Universidad de Zaragoza, Pedro Cerbuna 12,  E-50009 Zaragoza, Spain

\vspace*{0.1cm}

$^b$Centro de Astropart{\'\i}culas y F{\'\i}sica de Altas Energ{\'\i}as (CAPA), 
Universidad de Zaragoza, Zaragoza, Spain

\vspace*{0.1cm}

$^c$Deptartament de F{\'\i}sica Qu{\`a}ntica i Astrof{\'\i}sica and Institut de Ci{\`e}ncies del Cosmos (ICCUB),\\
Universitat de Barcelona, Mart{\'\i} i Franqu{\`e}s 1, E-08028 Barcelona, Catalonia, Spain

}

\end{center}

\vspace*{0.1cm}

\begin{abstract}
\noindent
The experimental measurements on flavour physics, in tension with
Standard Model predictions, exhibit large sources of Lepton Flavour
Universality violation. This note summarises an analysis of the effects of the
global fits to the Wilson coefficients assuming a model independent
effective Hamiltonian approach, by including a proposal of different
scenarios to include the New Physics contributions. Additionally, we
include an overview of the impact of the future generation of
colliders in the field of $B$-meson anomalies. 

\end{abstract}

\vspace*{0.5cm}
{\noindent Talk presented at the International Workshop on Future Linear Colliders (LCWS2021), 15-18 March 2021. C21-03-15.1.}

\def\thefootnote{\arabic{footnote}}
\setcounter{page}{0}
\setcounter{footnote}{0}

\newpage

\section{Introduction}

In the last few years, 
several experimental collaborations observed Lepton Flavour Universality Violating (LFUV) processes in
$B$ meson decays that would be a clear sign for physics beyond the Standard
Model (SM). In the $b\to c \ell \nu$ transitions, signs of violation of lepton
universality have been observed in the $e-\tau$, $\mu-\tau$ and $e-\mu$ 
cases~\cite{Abdesselam:2017kjf,Jung:2018lfu,Bobeth:2021lya}. The $\RDp^\ell$ and $\RDp^\mu$ ratios, defined by, 
\begin{equation}
\RDp^\ell = \frac{\mathrm{BR}(B \to D^{(*)} \tau \bar{\nu}_\tau ) }{[\mathrm{BR}(B \to D^{(*)} e \bar{\nu}_e) + \mathrm{BR}(B \to D^{(*)} \mu \bar{\nu}_\mu)]/2}\ ,
\end{equation}
and
\begin{equation}
\RDp^\mu = \frac{\mathrm{BR}(B \to D^{(*)} \tau \bar{\nu}_\tau ) }{ \mathrm{BR}(B \to D^{(*)} \mu \bar{\nu}_\mu)}\ ,
\end{equation}
have received special attention. 
The measurements of these ratios at BaBar~\cite{Lees:2012xj}, Belle~\cite{Abdesselam:2019dgh} and 
LHCb~\cite{Aaij:2017uff} experiments are larger than the SM prediction ($R_D^{\ell\ \mathrm{SM}} = 0.299
\pm 0.003$, $R_{D^*}^{\ell\ \mathrm{SM}} = R_{D^*}^{\mu\ \mathrm{SM}} =
0.258 \pm 0.005$~\cite{Amhis:2019ckw}). The world average of the 
experimental values for the \RDp ratios, as obtained by the Heavy
Flavour Averaging Group (HFLAV), assuming universality in the lighter 
leptons, is~\cite{Amhis:2019ckw}
\begin{equation}
R_D^\mathrm{ave} = 0.340 \pm 0.027 \pm 0.013,\qquad\qquad
R_{D^*}^\mathrm{ave} = 0.295 \pm 0.011 \pm 0.008.
\end{equation}
$R_D$ exceeds the SM value by $1.4\,\sigma$, and $R_{D^*}$ by $2.5\,\sigma$. When combined together, included their correlation, the excess is $3.08\,\sigma$.

Another class of $B$ meson observables showing signs of LFUV is related
to $b  \to s \ell^+ \ell^- $ processes, namely the optimised angular
observable $P_5'$ \cite{DescotesGenon:2012zf} and the \RKp ratios,
\begin{equation}
\RKp = \frac{\mathrm{BR}(B\to K^{(*)} \mu^+ \mu^- )}{\mathrm{BR}(B\to K^{(*)} e^+ e^- )}\ .
\end{equation}
As a consequence of Lepton Flavour Universality (LFU),
$R_K = R_{K^*} = 1$ with uncertainties of the order of $1\%$ in the
SM~\cite{Hiller:2003js,Bordone:2016gaq}. 
These ratios are observables that have small theoretical
uncertainties. The latest experimental results from
LHCb, in the specified regions of $q^2$ di-lepton invariant mass, are:
\begin{align}
R_K^{[1.1, 6]} = 0.846^{+0.042}_{-0.039}{}^{+0.013}_{-0.012}\, \qquad &\mbox{\cite{Aaij:2021vac} }\nonumber \\
R_{K^*}^{[0.045,1.1]} = 0.66^{+0.11}_{-0.07}\pm 0.03 \qquad\qquad
R_{K^*}^{[1.1,6]} = 0.69^{+0.11}_{-0.07}\pm 0.05\ . \qquad
&\mbox{\cite{Aaij:2017vbb} }
\end{align}
The compatibility of the individual measurements with respect to the SM
predictions is of $3.1\, \sigma$ for the $R_K$ ratio, $2.3\, \sigma$ for the
$R_{K^*}$ ratio in the low-$q^2$ region and $2.4\, \sigma$ in the
central-$q^2$ region. The Belle collaboration has also recently reported
experimental results for the \RKp ratios~\cite{Abdesselam:2019lab,Abdesselam:2019wac}, 
although with less precision than the LHCb measurements. 

A great theoretical effort has been devoted to the understanding of the
deviations in the $\RKp$ and $\RDp$ observables, 
and combined explanations for those deviations
(see, for example~\cite{Altmannshofer:2017fio,Hiller:2014ula,Hurth:2016fbr,Altmannshofer:2017yso,Geng:2017svp,Ciuchini:2017mik,Alda:2018mfy,Alok:2017qsi,Jung:2018lfu,Bhattacharya:2018kig,Murgui:2019czp,Blanke:2019qrx,Bhattacharya:2014wla,Calibbi:2015kma,Hiller:2016kry,Bhattacharya:2016mcc,Crivellin:2017zlb,Cai:2017wry,Alok:2017jaf,Feruglio:2017rjo,Buttazzo:2017ixm,DiLuzio:2017vat,Bordone:2017bld,Kumar:2018kmr,Angelescu:2018tyl,Crivellin:2018yvo,Bifani:2018zmi,Babu:2020hun,Altmannshofer:2020axr}
and references therein). Besides, the experimental data has been used to
constrain New Physics (NP) models. 
Several global fits have been performed in the
literature~\cite{Capdevila:2017bsm,Celis:2017doq,Alok:2017sui,Camargo-Molina:2018cwu,Datta:2019zca,Aebischer:2019mlg,Aoude:2020dwv,Alda:2020okk}. 

These proceedings are mainly based on our previous work in~\cite{Alda:2018mfy,Alda:2020okk} 
where we investigate the effects of the global fits to the Wilson
coefficients assuming a model independent effective Hamiltonian
approach. In section~\ref{sec:EFT} we present a brief summary of the Effective Field
Theory used to describe possible NP contributions to $B$ decays 
observables. A summary of the results obtained in~\cite{Alda:2018mfy} 
for a fit of the \RKp ratios and the
angular observables $P'_4$ and $P'_5$ to the Weak Effective Theory
Wilson coefficients is included in this section. 
Section~\ref{sec:fits} is devoted to the global fits to the Wilson
coefficients, presenting the set of scenarios that we have defined 
in~\cite{Alda:2020okk} for the
phenomenological study, by considering the NP contributions to the Wilson
coefficients in such a way that NP is present in one, two or three of the Wilson
coefficients simultaneously. These scenarios are used to study the
impact of the global fits to the Wilson coefficients and, therefore, 
to exhibit more clearly which combinations of Wilson coefficients are 
preferred and/or constrained by experimental data. 
We complement our results with a discussion in
section~\ref{sec:colliders} of the impact that future $e^+ e^-$ 
linear colliders will have in the $B$ anomalies~\cite{Alda:2021}. 
Conclusions are presented in section~\ref{sec:conclu}.

\section{Effective field theories for $B$ observables}\label{sec:EFT}

One of the most widely used tools to study any possible New Physics (NP)
contribution is the Effective Field Theory. The effective Hamiltonian approach allows us to
perform a model-independent analysis of NP effects. In this way, it is possible to
obtain constraints on NP contributions to the Wilson coefficients of the
Hamiltonian from the experimental results. 

The Standard Model
Effective Field Theory (SMEFT) is formulated at an energy scale 
$\mu_\mathrm{SMEFT} = \Lambda$ higher that the electroweak (EW) scale, and
the degrees of freedom are all SM fields. The Weak Effective Theory
(WET) is formulated at an energy scale below the EW scale, 
for example $\mu_\mathrm{WET} = m_b$, and the top quark, Higgs, $W$ and $Z$ bosons are integrated out. 

The relevant terms of the WET Lagrangian \cite{Buras:1998raa,Aebischer:2015fzz,Tanaka:2012nw} are:
\begin{equation}
\mathcal{L}_{\text{eff}} = -\frac{4 G_F}{\sqrt{2}}V_{cb}\sum_{\ell = e, \mu, \tau} (1 + C_{VL}^\ell) \mathcal{O}_{VL}^\ell + \frac{4G_F}{\sqrt{2}}V_{tb}V_{ts}^*\frac{e^2}{16\pi^2}\sum_{\ell=e,\mu} (C_9^\ell \mathcal{O}_9^\ell  + C_{10}^\ell \mathcal{O}_{10}^\ell) \ ,
\end{equation}
where $G_F$ is the Fermi constant, $e$ is the electromagnetic coupling, $V_{qq'}$
are the elements of the Cabibbo-Kobayashi-Maskawa (CKM) matrix and with the dimension six
operators defined as,  
\begin{equation}
\mathcal{O}_{VL}^\ell = (\bar{c}_L \gamma_\alpha b_L)(\bar{\ell}_L \gamma^\alpha \nu_\ell)\ ,\qquad \mathcal{O}_9^\ell = (\bar{s}_L \gamma_\alpha b_L)(\bar{\ell} \gamma^\alpha \ell)\ , \qquad \mathcal{O}_{10}^\ell = (\bar{s}_L \gamma_\alpha b_L)(\bar{\ell} \gamma^\alpha \gamma_5 \ell) \ ,
\end{equation}
and their corresponding Wilson coefficients $C_{VL}^\ell$, $C_9^\ell$
and $C_{10}^\ell$. The $C_9^\ell$ and $C_{10}^\ell$ Wilson coefficients
have contributions from the SM processes as well as any NP contribution, 
\begin{equation}
C_i^\ell = C_i^{\mathrm{SM}\, \ell} + C_i^{\mathrm{NP}\, \ell}\ ,\qquad\qquad i= 9,10\ .
\end{equation}

The dependence of the \RDp ratios on the Wilson coefficients is given 
by~\cite{Bhattacharya:2016mcc,Feruglio:2017rjo}:
\begin{align}
\RDp^\ell &= \RDp^{\ell, \mathrm{SM}} \frac{|1+ C_{VL}^\tau|^2}{ (|1+C_{VL}^e|^2 + |1+C_{VL}^\mu|^2)/2}\ , \nonumber\\
\RDp^\mu &= \RDp^{\mu, \mathrm{SM}} \frac{|1+ C_{VL}^\tau|^2}{ |1+C_{VL}^\mu|^2}\ . \label{eq:RD}
\end{align}

For the \RKp ratios, the dependence on the Wilson coefficients has been
previously obtained in \cite{Alda:2018mfy}, where an analytic 
computation of $\Rks$ as a function of $\CnmuNP$, $\CtmuNP$ in the
region $1.1 \leq q^2 \leq 6.0\GeV^2$
was performed. The result is given by~\cite{Alda:2018mfy}: 
\begin{equation}
R_{K^*}^{[1.1,6]} \simeq \frac{0.9875+0.1759\, \mathrm{Re}\,\CnmuNP - 0.2954\,  \mathrm{Re}\, \CtmuNP + 0.0212|\CnmuNP|^2 + 0.0350 |\CtmuNP|^2}{1\,\  \ \ \ \ +0.1760\, \mathrm{Re}\,\CneNP - 0.3013\, \mathrm{Re}\,  \CteNP + 0.0212|\CneNP|^2 + 0.0357 |\CteNP|^2}\ .
\end{equation}

\begin{figure}
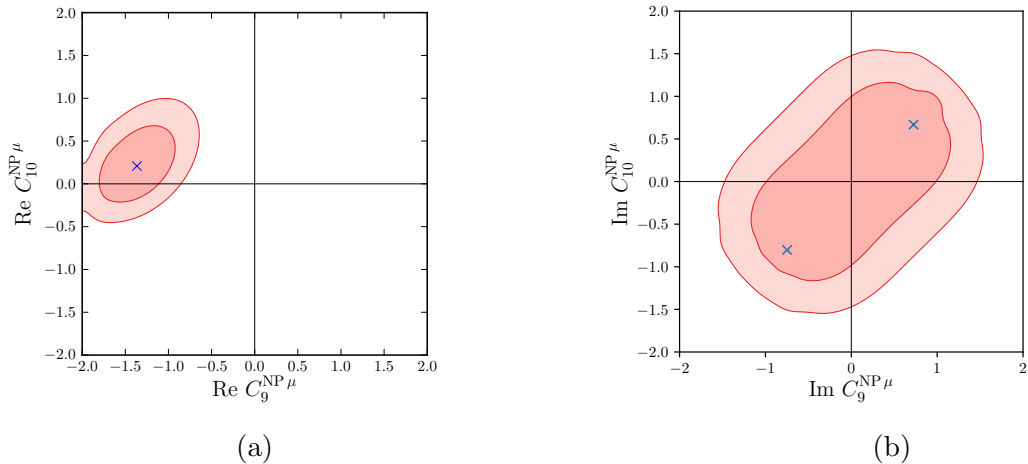

\centering
\begin{tabular}{cc}
\resizebox{0.5\textwidth}{!}{\input{fitre.pgf}}&\hspace*{-1.2cm}
\resizebox{0.5\textwidth}{!}{\input{fitIm_C9C10.pgf}}\\
(a) & (b)
\end{tabular}
\caption{Best fit and $1\sigma$ and $2\sigma$ contours to semi-leptonic
  $B$-decays observables, $R_K$, $R_{K^{*0}}$, $P'_4$ and $P'_5$, using
  (a) real and (b) imaginary Wilson coefficients.}\label{im:WETfits}
\end{figure}

In~\cite{Alda:2018mfy} we performed a fit of the \RKp ratios and the
angular observables $P'_4$ and $P'_5$ to the WET Wilson coefficients
\CnmuNP and \CtmuNP. We considered two hypothesis: both coefficients
being real numbers or imaginary numbers. The allowed regions at
$1\, \sigma$ and $2\, \sigma$ are shown in Figure~\ref{im:WETfits}. The best
fit to the real coefficients is located at $\CnmuNP = -1.09$, $\CtmuNP =
0.48$ improves the SM predictions by $5.95\,\sigma$, showing a clear
preference for non-zero NP contribution to $C_9^\mu$. The imaginary fit
presents two nearly symmetric minima located at $\CnmuNP = -0.75\,i$,
$\CtmuNP = -0.74\,i$ and $\CnmuNP = 0.72\,i$, $\CtmuNP=0.74\,i$, with a
pull from the SM of $0.9\,\sigma$. In conclusion, purely imaginary
Wilson coefficients do not provide a good description of the data. 
Therefore we will only consider real Wilson coefficients in what follows.

The NP contributions at an energy scale $\Lambda$ ($\Lambda \sim
\mathcal{O}(\mathrm{TeV})$) is described by the SMEFT
Lagrangian as given in~\cite{Grzadkowski:2010es},
\begin{equation}
  \mathcal{L}_\mathrm{SMEFT} =
  \frac{1}{\Lambda^2}\left(\Clqo^{ijkl}\, O_{\ell q(1)}^{ijkl} +
    \Clqt^{ijkl}\,  O^{ijkl}_{\ell q(3)}   \right) \ ,
\label{eq:Lagr_SMEFT}
\end{equation}
where the dimension six operators are defined as 
\begin{equation}
O_{\ell q(1)}^{ijkl} = (\bar{\ell}_i \gamma_\mu P_L \ell_j)(\bar{q}_k \gamma^\mu  P_L q_l),\qquad\qquad O_{\ell q(3)}^{ijkl}= (\bar{\ell}_i \gamma_\mu \tau^I P_L \ell_j)(\bar{q}_k \gamma^\mu \tau^I P_L q_l) ,
\end{equation}
 $\ell$ and $q$ are the lepton and quark $SU(2)_L$ doublets, $\tau^I$ the Pauli matrices, and ${i,j,k,l}$ denote generation indices. The $O_{\ell q(1)}$ operator couples two $SU(2)_L$-singlet currents, while the $O_{\ell q(3)}$ operator couples two $SU(2)_L$-triplet currents. Consequently, $O_{\ell q(1)}$ only mediates flavour-changing neutral processes, and $O_{\ell q(3)}$ mediates both flavour-changing neutral and charged processes. 
We will restrict our analysis to operators including only third generation quarks and 
same-generation leptons, and we will use the following notation for their Wilson coefficients: 
\begin{equation}
  \Clq^e \equiv \Clq^{1133}\ , \qquad\qquad \Clq^\mu \equiv \Clq^{2233}\ ,
  \qquad\qquad \Clq^\tau \equiv \Clq^{3333}\ .
\label{eq:wcs}
\end{equation}
This particular choice of the Wilson coefficients
is motivated by the fact that the most prominent discrepancies
between SM predictions and experimental measurements, namely
$R_{K^{(*)}}$ and $R_{D^{(*)}}$, affect the third quark generation. From
a symmetry point of view, this would amount to imposing an 
$U(2)^3 = U(2)_q \times U(2)_u \times U(2)_d$ symmetry between the first
and second quark generations~\cite{Barbieri:2011ci,Barbieri:2012uh,AguilarSaavedra:2018nen}, 
that remain SM-like. In the lepton sector we only consider diagonal entries in 
order to avoid Lepton Flavour Violating (LFV) decays.

These operators generate the $C_{VL}^\ell$, $C_9^\ell$ and $C_{10}^\ell$
operators of the electroweak effective field theory when matched at the EW 
scale $\mu_\mathrm{EW}$. Using the package \texttt{wilson}
\cite{Aebischer:2018bkb}, we define the $\Clq$ operators at $\Lambda = 1
\mathrm{TeV}$, we calculate their running down to $\mu_\mathrm{EW} =
M_Z$, then match them with the EW operators and finally run the
down to $\mu = m_b$, where the $B$-physics observables are computed. We
found the following relations between the Wilson coefficients at high
and low energies: 
\begin{align}
C_9^{\mathrm{NP}\ e, \mu} = -0.583 \, \Clqo^{e, \mu} - 0.596 \, \Clqt^{e,
  \mu}\ , &\qquad\qquad C_{10}^{\mathrm{NP}\ e, \mu} = 0.588 \, \Clqo^{e,
  \mu} + 0.591 \, \Clqt^{e, \mu}\ , \nonumber\\
C_{VL}^{e, \mu} = 0.0012 \, \Clqo^{e, \mu} - 0.0644\, \Clqt^{e, \mu}\ ,
          &\qquad\qquad C_{VL}^\tau = -0.0598\, \Clqt^\tau\ .
\label{eq:running}
\end{align}

The $\mathcal{O}_{\ell q}$ operators also produce unwanted contributions
to the $B \to K^{(*)} \nu \bar{\nu}$
decays~\cite{Feruglio:2017rjo,Aebischer:2018iyb}.
In order to obey these constraints, we will fix the relation

\begin{equation}
\Clqo^i = \Clqt^i \equiv \Clq^i\ .
\end{equation}
This relation also has the positive consequence of a partial
cancellation of loop-induced effects in $Z$-pole and LFV observables.

\section{Global fits}\label{sec:fits}

The effective operators affect a large number of observables. Therefore, 
any NP prediction based on Wilson coefficients has to be confronted not 
only with the \RKp an \RDp measurements, but also with additional
several measurements involving the decays of $B$ mesons. In the case of 
the SMEFT, the Renormalization Group evolution produces a
mix of the low-energy effective operators and then, 
modifies the $W$ and $Z$ couplings to leptons. In consequence, NP in
the top sector will indirectly affect EW observables, such as
the mass of the $W$ boson, the hadronic cross-section of the $Z$ boson 
$\sigma^0_\mathrm{had}$ or the branching ratios of the $Z$ to different
leptons. In order to keep the predictions consistent with this range of 
experimental test, global fits have proven to be a valuable tool~\cite{Celis:2017doq,Camargo-Molina:2018cwu,Aebischer:2019mlg,Aoude:2020dwv}. 

In~\cite{Alda:2020okk} we have performed global fits to the $\Clq$ Wilson coefficients using
the package \texttt{smelli} v1.3~\cite{Aebischer:2018iyb}. The global fit
includes the \RKp and \RDp observables, the $W$ and $Z$ decay widths,
the branching ratios to leptons,
the $b\to s \mu\mu$ observables (including $P_5'$ and the branching ratio of
$B_s \to \mu\mu$) and the $b \to s \nu \bar{\nu}$ observables. The SM input parameters
used in the analysis are explicitely given in~\cite{Alda:2020okk}. They are
taken from open source code \texttt{flavio} v1.5~\cite{Straub:2018kue}, sources used by the
program are quoted when available. Concretely, we have supplemented the experimental
measurements of the \texttt{flavio} v1.5 database with updated 
values for $R_K$~\cite{Aaij:2021vac},  $\RDp$~\cite{Abdesselam:2019lab},
$B\to K^*\ell^+\ell^-$ differential
observables~\cite{Aaij:2020nrf,Aaij:2020umj}, $B_{(s)}\to\mu^+\mu^-$~\cite{LHCb:2020zud} 
and a re-analysis of the EW precision tests from LEP~\cite{Janot:2019oyi}. 

We have defined some specific scenarios, shown in Table~\ref{tab:Fits}, 
for combinations of the $\Clq^i$ operators such that NP contributions
to the Wilson coefficients emerge in one, two or three of the Wilson 
coefficients simultaneously~\cite{Alda:2020okk}: in Scenarios I-III NP only modifies the
$\Clq$ operators in one lepton flavour at a time; in Scenarios IV-VI 
NP is present in two of the Wilson coefficients simultaneously; and finally
in Scenarios VII-IX we consider the more general case in which three of
the $\Clq^i$ operators receive NP contributions. The more general one of
these last three scenarios is Scenario VII, in which we consider three 
independent Wilson coefficients.

The goodness of each fit is evaluated with its difference of $\chi^2$
with respect to the SM, $\Delta \chi^2_\mathrm{SM} = \chi^2_\mathrm{SM} -
\chi^2_\mathrm{fit}$. The package \texttt{smelli} actually computes the
differences of the logarithms of the likelihood function 
$\Delta \log L = -\frac{1}{2} \Delta \chi^2$. In order to compare two 
fits $A$ and $B$, we use the pull between them in units of $\sigma$, 
defined as \cite{Descotes-Genon:2015uva,Capdevila:2018jhy}
\begin{equation}
\mathrm{Pull}_{A \to B} = \sqrt{2} \mathrm{Erf}^{-1}[F(\Delta \chi^2_A - \Delta \chi^2_B; n_B - n_A )]\,,
\end{equation}
where $\mathrm{Erf}^{-1}$ is the inverse of the error function, $F$ is
the cumulative distribution function of the $\chi^2$ distribution and
$n$ is the number of degrees of freedom of each fit. We will compare
each scenario against two cases: the SM ($\Clq = 0$,
$n=0$) and the best fit point using three independent Wilson coefficients (scenario
VII). The pull from the
SM quantifies how much each scenario is preferred over the SM to
describe the data. The larger the pull, the better description of
the data of the preferred scenario. The pull of scenario VII quantifies
how much the fit over the whole space of parameters is preferred over the simpler and
more constrained fits. From the analysis of this pull we are able to discuss the
relevance of the proposed scenarios, the larger the pull means that the more
restricted scenario represents a worser description of the experimental data.

\begin{table}
\centering
\begin{tabular}{|c|c|c|c|c|c|c|c|}\hline
\multicolumn{2}{|c|}{\multirow{2}*{Scenario}}&\multirow{2}*{$C_{\ell
    q}^e$} &\multirow{2}*{$C_{\ell q}^\mu$}&\multirow{2}*{$C_{\ell
    q}^\tau$} &\multirow{2}*{$\Delta\chi^2_\mathrm{SM}$} & Pull & Pull
\\ \multicolumn{2}{|c|}{} & & & & & from SM & to VII\\\hline
I& $e$	& $-0.14 \pm 0.04$ & & & 8.84 &	2.97 $\sigma$ & 4.37 $\sigma$\\\hline
II& $\mu$ & & $0.10 \pm 0.04$ & & 5.47 & 2.34 $\sigma$ & 4.73 $\sigma$\\\hline
III& $\tau$ & & & $-0.38\pm0.19$ & 3.85 & 1.96	$\sigma$ & 4.89 $\sigma$\\\hline
IV& $e$ and $\mu$ & $-0.25 \pm 0.07$ &	$0.24 \pm 0.06$	& & 28.42 & 4.97
$\sigma$ & 1.75 $\sigma$\\\hline
V& $e$ and $\tau$ & $-0.14 \pm 0.06$ & & $-0.4 \pm 0.3$ & 12.98 & 3.17 $\sigma$ & 4.30 $\sigma$\\\hline
VI& $\mu$ and $\tau$ &	& $0.10 \pm 0.06$ & $-0.3 \pm 0.3$ & 8.73 & 2.49
$\sigma$ & 4.77 $\sigma$\\\hline
VII& $e$, $\mu$ and $\tau$ & $-0.25 \pm 0.02$ & $0.211 \pm 0.016$ &
$-0.3 \pm 0.4$ & 31.50 & 4.97 $\sigma$ &\\\hline
VIII& $e = \mu = \tau$	& $-0.0139 \pm 0.0003$ & $-0.0139 \pm 0.0003$ &
$-0.0139 \pm 0.0003$ & 0.30 & 0.55 $\sigma$ & 5.23 $\sigma$\\\hline
IX& $e = -\mu = \tau$ & $-0.232\pm0.001$ & $0.232\pm0.001$ &
$-0.232\pm0.001$ & 30.74 & 5.54 $\sigma$ & 0.41 $\sigma$\\\hline
\end{tabular}
\caption{Best fit values and pulls from the Standard Model and of
  scenario VII for several combinations of $\Clq^i$ operators.}
\label{tab:Fits}
\end{table}

The results of the fits are summarised in Table~\ref{tab:Fits} for
several combinations of $\Clq^i$ operators, with one, two or three lepton flavour
present simultaneously in the Wilson coefficients.
The best fit values at 1 $\sigma$ and pulls from the SM and to scenario
VII for all cases are included in this table.

Summarising the results, we found that the largest pull from the SM 
prediction when NP only modifies the $\Clq^i$ operators in one lepton flavour at a time, i.e
  $\Clq^e$, $\Clq^\mu$ or $\Clq^\tau$, is obtained in scenario I 
where the coupling to electrons is added. It is almost $3\ \sigma$. 
This result is a reflection of the great impact of the EW precision 
observables in the global fit. If we restricted our fit to only 
$b \to s \ell^+ \ell^-$ observables, the fit to only muons in scenario
II would display a better pull from the SM of $2.34\ \sigma$, in 
line with the common wisdom about the anomalies, explaining them through
NP in the muon
sector~\cite{Altmannshofer:2017yso,Ciuchini:2017mik,Capdevila:2017bsm,Descotes-Genon:2015uva,DAmico:2017mtc}. The
worst pull is obtained in the fit to the tau coefficient, 
with $1.96\ \sigma$, as it does not modify the value of the \RKp
ratios. On the other hand, scenarios I and II both produce SM-like predictions for the observables 
$R_D$ and $R_{D^*}$. Scenario III, with a larger value of its Wilson
coefficient, produces values closer to the experimental measurements;
i.e $R^\ell_D = 0.318$ and $R^\ell_{D^*}=0.268$. In order to fully
address the anomaly in these observables, a larger deviation from the 
SM would be needed; however such a deviation would be in conflict with 
the EW precision data, as we obtained in~\cite{Alda:2020okk}, 
and in agreement with~\cite{Capdevila:2017iqn}.

\begin{figure}
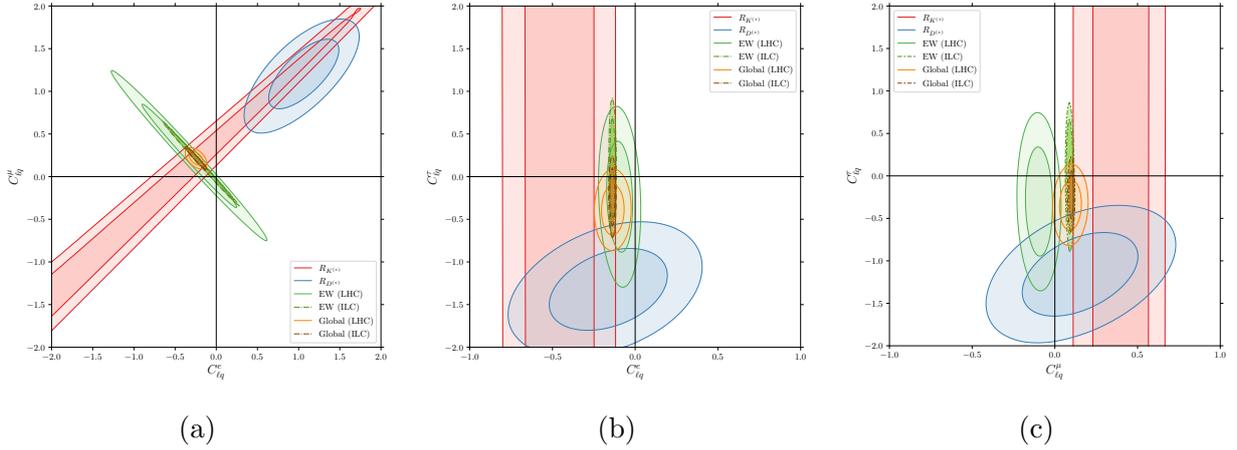

\begin{center}
\begin{tabular}{ccc}
\resizebox{0.3\textwidth}{!}{\input{scIV_EW.pgf}}&
\resizebox{0.3\textwidth}{!}{\input{scV_EW.pgf}}&
\resizebox{0.3\textwidth}{!}{\input{scVI_EW.pgf}}\\
(a)&(b)&(c)
\end{tabular}
\caption{$1\sigma$ and $2\sigma$ contours for scenarios with two
  lepton flavours present 
in the Wilson coefficients: (a) Scenario IV, (b) Scenario V, and (c)
Scenario VI. All available data are considered.}
\label{im:globalfits}
\end{center}
\end{figure}

For the scenario in which NP is present in two of the Wilson
coefficients, the best fit corresponds to scenario IV, where the 
contributions to $\Clq^e$ and $\Clq^\mu$ are favoured with a pull 
of $4.97\ \sigma$ with respect to the SM. Figure~\ref{im:globalfits} 
shows the allowed regions for these fits. In the fit to Scenario IV, 
the \RKp and \RDp observables constrain the $\Clq^e - \Clq^\mu$ combination; while the
  LFU-conserving EW precision observables tightly constrain the 
combination $\Clq^e + \Clq^\mu$. Clearly, the EW precision
observables play an important role in the global fit and the preferred
values for the Wilson coefficients. The reason for this behaviour is
justified by deviations in $Z$-couplings to leptons, the $\tau$-leptonic
decays and the Z and W decays widths, as shown in~\cite{Feruglio:2018jnu}. 
The values of the \RKp and \RDp observables in this scenario are given in 
Table~\ref{tab:observables}. Together, these
  sets of observables constrain the fit to a narrow ellipse around the
  best fit point. In Scenarios V and VI, the $\Clq^\tau$ coefficient is
  determined by the EW precision observables, that are
  compatible with a SM-like coefficient, and by \RDp observables, that
  prefer a large negative value. All the experimental constraints for
  $\Clq^\tau$ show large uncertainties, which result in less statistical
  significance of these fits and $\Clq^\tau$ still being compatible with
  zero at $2\,\sigma$ level.

\begin{figure}
\begin{center}
\begin{tabular}{cc}
\includegraphics[width=0.45\textwidth]{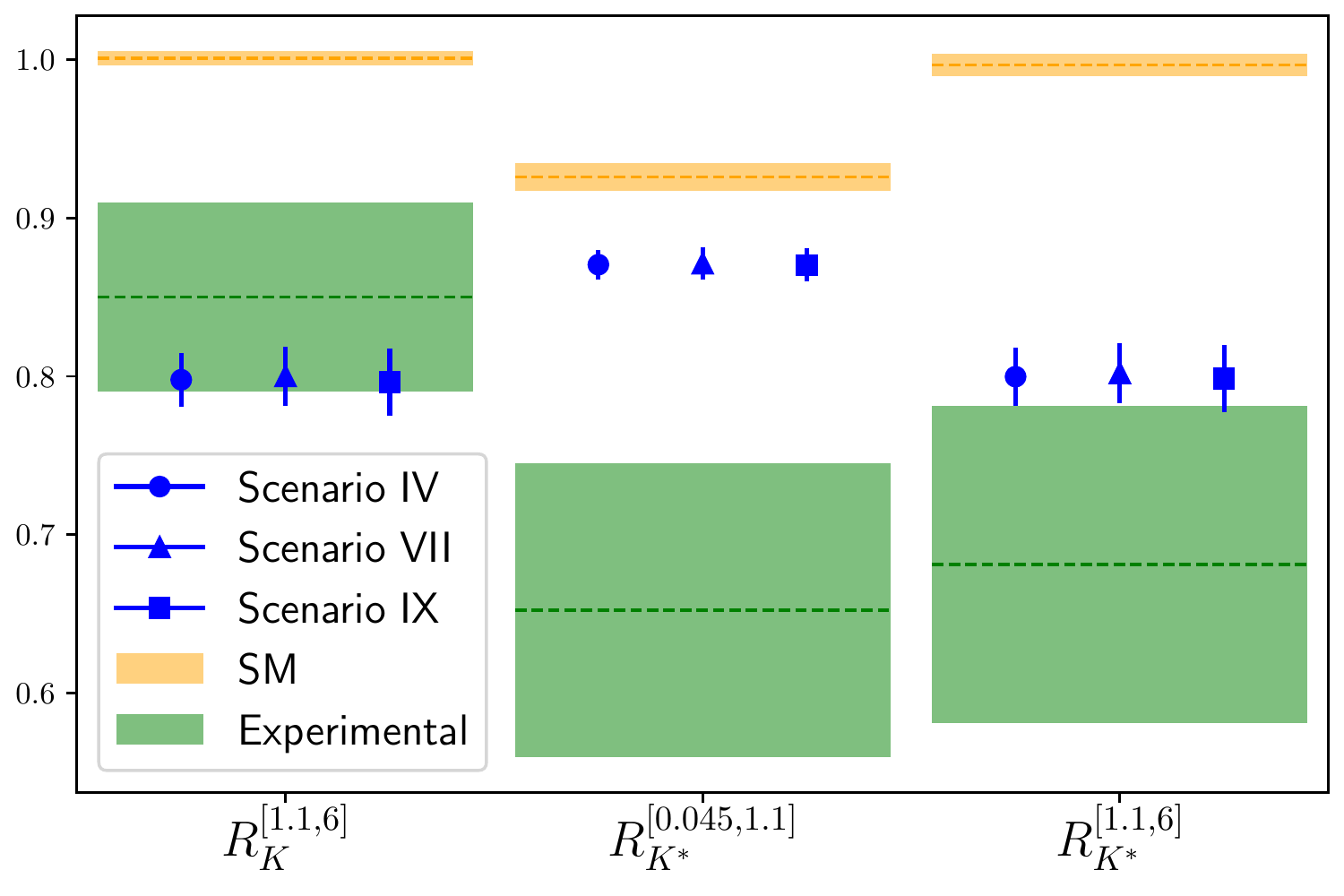}&
\includegraphics[width=0.45\textwidth]{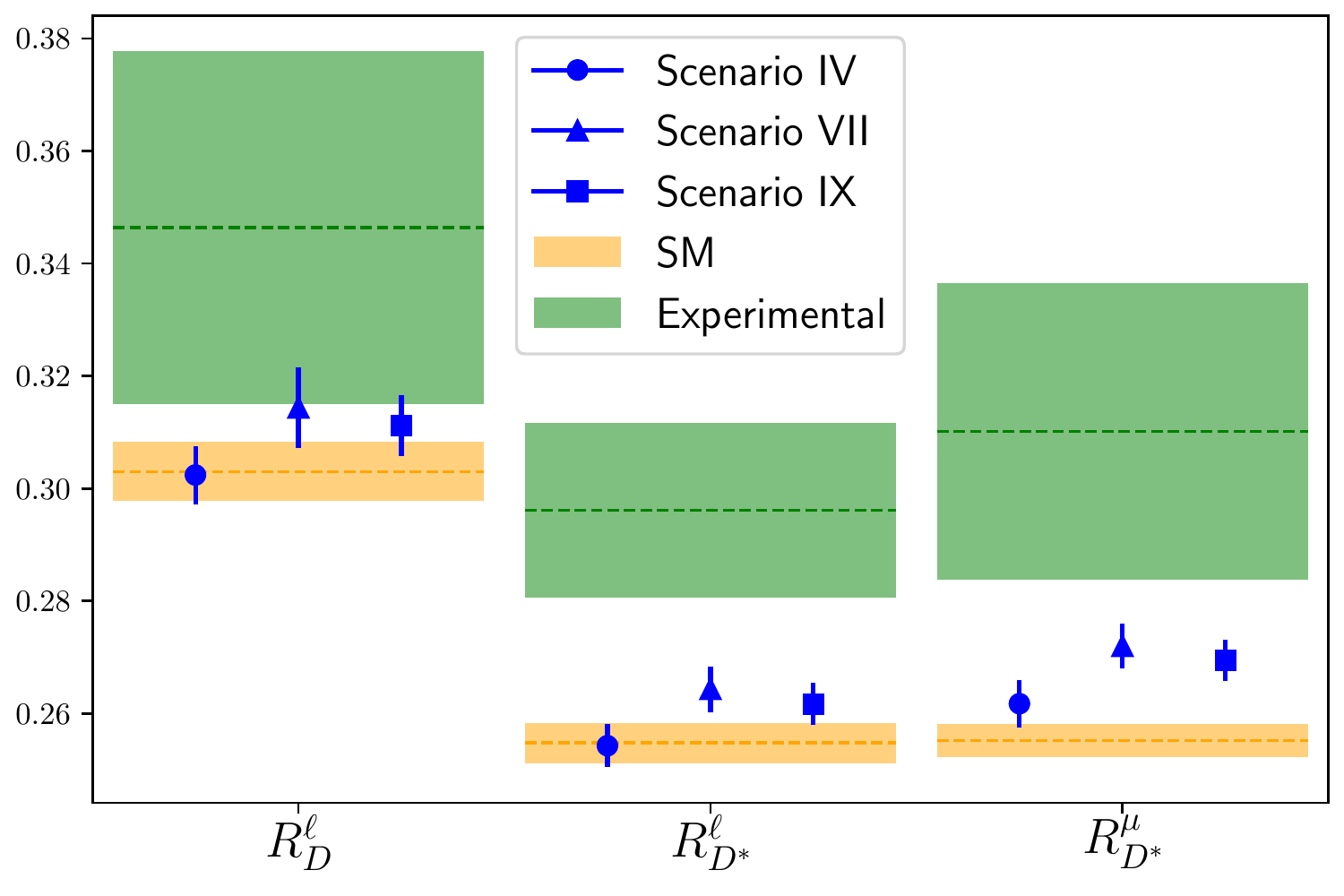}\vspace*{-0.3cm}\\
(a)&(b)
\end{tabular}
\caption{Central value and $1 \sigma$ uncertainty of the (a) $\RKp$
  observables, and (b) $\RDp$ observables (blue lines) in scenarios 
IV, VII and IX, compared to the SM prediction (yellow) and experimental measurements (green).}
\label{im:RK}
\end{center}
\end{figure}

As already established, the more general cases are the ones in which three of
the $\Clq^i$ operators receive NP contributions. A particular scenario
corresponds with \textit{universal} couplings (Scenario VIII); 
i.e the three Wilson coefficients have the same
  \textit{universal} contribution, and does not violate LFU. We found
  the smallest pull with respect to the SM ($0.55\,\sigma$) in this
  case, which shows that LFU NP can not explain experimental data and,
  therefore, LFU violation is needed to accommodate it.
When the three $\Clq$ operators receive independent NP contribution
(Scenario VII), the pull from the SM, $4.97\ \sigma$, is
  similar to that of scenario IV, and the values of $\Clq^e$ and
  $\Clq^\mu$ are similar too, therefore the predictions for the \RKp
  observables are very similar, as shown in Figure~\ref{im:RK}a. The
  value of $\Clq^\tau$ is close to that of Scenarios III, V and VI, which
  allows a best fit to the \RDp observables, and especially to
  $R_D^\ell$, that is compatible at $1\,\sigma$ with its experimental
  value, as shown in Figure~\ref{im:RK}b. Therefore, we conclude that
  the prediction of the $\RDp$ observables is improved in scenario VII.
  This scenario was analysed in more detail in~\cite{Alda:2020okk}.
  We found that the constraints to the fit can be explained by the
  combined effect of three different classes of observables: in the first
  place, the linear combination $C_3 \sim (-\Clq^e + \Clq^\mu)/\sqrt{2}$
  shows a clear preference for a LFU-violating situation, driven mostly
  by $R_K$ and $R_{K^*}$, and in tension with $\mathrm{BR}(\pi^+\to e\nu)$,
  $R_{D^*}^{e/\mu}$ and $R_{e\mu}(K^+\to \ell^+\nu)$. The second class of
  observables are LFU-conserving, affecting the linear combination
  $C_2 \sim (-\Clq^e - \Clq^\mu)/\sqrt{2}$, the more relevant observables
  being the EW precision tests (the mass of the $W$ boson $m_W$,
  the $Z$-decay asymmetries $A_e$, $A_\tau$ and $A_\mathrm{FB}$ and the
  $Z$ decay width $\Gamma_Z$). Our fit is less sensitive to the third class
  of observables, those that affect $\tau$ physics in $C_1 \sim - \Clq^\tau$,
  where the more relevant constraints come from the leptonic decays
  $\tau \to e \bar{\nu}\nu$ and $\tau \to \mu \bar{\nu}\nu$, the hadronic
  cross-section of the $Z$ $\sigma^0_\mathrm{had}$, and the ratios $\RDp$.
  The LFU-violating observables, as well as the $\tau \to e \bar{\nu}\nu$ decay,
  proved to be the most relevant observables in the fit overall. Finally,
  Scenario IX corresponds with the three Wilson
  coefficients having the same absolute value, but $\Clq^\mu$ has 
the opposite sign. This particular arrangement of the coefficients 
was inspired by the similar absolute values of $\Clq^e$ and $\Clq^\mu$ 
in Scenario VII. This choice produces a good fit, with a pull of 
$5.54\,\sigma$. It is also the only scenario that remains compatible at $1\,\sigma$ with scenario VII. 

\begin{table}
\centering
\begin{tabular}{|c|c|c|c|c|}\hline
Observable & Scenario IV & Scenario VII & Scenario IX & Measurement \\\hline
$R_K^{[1.1, 6]}$ & $0.799 \pm 0.017$ & $0.800 \pm 0.018$ & $0.79 \pm 0.02$ & $0.85 \pm 0.04$ \\\hline
$R_{K^*}^{[0.045,\ 1.1]}$ & $0.870 \pm 0.009$ & $0.871 \pm 0.010$ & $0.870 \pm 0.010$ & $0.65 \pm 0.09$\\\hline
$R_{K^*}^{[1.1,\ 6]}$ & $0.800 \pm 0.018$ & $0.802 \pm 0.019$  & $0.80 \pm 0.02$ & $0.68 \pm 0.10$ \\\hline
$R_D^\ell$ & $0.302 \pm 0.005$ & $0.314 \pm 0.007$ & $0.311 \pm 0.005$ & $0.35 \pm 0.03$ \\\hline
$R_{D^*}^\ell$ & $0.254 \pm 0.004$ & $0.264 \pm 0.004$ & $0.261 \pm 0.004$ & $0.296 \pm 0.016$ \\\hline
$R_{D^*}^\mu$ & $0.261 \pm 0.004$ & $0.272 \pm 0.004$ & $0.269 \pm 0.004$ & $0.31 \pm 0.03$ \\\hline
\end{tabular}
\caption{Values of the $\RKp$ and $\RDp$ observables in the scenarios with best pulls.}
\label{tab:observables}
\end{table}
The results for the \RKp and \RDp observables in the scenarios with
best pulls, Scenarios IV, VII and IX, are presented in 
Table~\ref{tab:observables}. For comparison, an statistical combination
of all the available measurements of each observable, performed by
\texttt{flavio} is included in the last column of this table. In the
case of the $\RDp$ ratios, this combination does not assume flavour universality
between electrons and muons. Figure~\ref{im:RK} shows the results 
for the central value and $1\,\sigma$ uncertainty of these two
observables in the three scenarios, compared to the SM prediction 
(yellow area) and experimental measurements (green area). These 
three scenarios have similar fits for the Wilson
coefficients $\Clq^e$ and $\Clq^\mu$, and therefore reproduce the 
experimental value of $R_K^{[1.1,6]}$ and reduce the tension in 
$R_{K^*}^{[1.1,6]}$. The main difference between Scenarios IV, VII and
IX is the fit for $\Clq^\tau$: Scenario IV has no NP contribution in 
the $\tau$ sector and consequently predicts SM-like \RDp ratios. 
Scenario VII has a large contribution to $\Clq^\tau$ and is able 
to produce a prediction for $R_D^\ell$ compatible with the experimental 
results, and significantly improve the predictions for $R_{D^*}^\ell$ 
and $R_{D^*}^\mu$. Scenario IX has an intermediate value of $\Clq^\tau$, 
and consequently its predictions for the \RDp ratios are not as good as in Scenario VII. 

In addition to the observables included in our global fits, it is also
possible to constrain the NP contributions to Wilson coefficients using 
high-energy collision data from
LHC~\cite{Faroughy:2016osc,Greljo:2018tzh}. We also have checked that all the results of
our fits are compatible with the limits imposed by the high-$p_T$ phenomena~\cite{Alda:2020okk}.

\section{Prospects from $e^+e^-$ colliders}\label{sec:colliders}

A new generation of particle colliders, complementary to the LHC and its
upgrade HL-LHC, will be ready in the coming decades. The International
Linear Collider (ILC) will be a linear $e^+ e^-$ collider in Japan,
operating at center-of-mass energies ranging from $\sqrt{s} =$250 GeV at 
the first stages up to $\sqrt{s} =$1 TeV \cite{Aihara:2019gcq}. The
Compact Linear Collider (CLIC) at CERN will also be a linear $e^+ e^-$
collider, operating from $\sqrt{s} =$ 380 GeV up to $\sqrt{s} =$ 3
TeV~\cite{Robson:2018enq}. The Future Circular Collider (FCC), also at
CERN, will be a circular collider first using electrons (FCC-ee) from 
$\sqrt{s} =$90 GeV ($Z$ pole) up to $\sqrt{s} =$365 GeV, and then using 
hadrons (FCC-hh) reaching $\sqrt{s} =$100 TeV \cite{Bordry:2018gri}. 
These colliders are conceived primarily as Higgs factories, exploring
the origin of the electroweak symmetry breaking mechanism and the 
hierarchy problem. But they can also supplement the flavour programs of 
the LHCb and Belle in different ways: by producing $b$-flavoured hadrons
in $e^+ e^- \to Z \to b\overline{b}$ events (ILC operating at the $Z$ pole is expected to 
produce around $10^9$ $Z$s (``GigaZ'')~\cite{Irles:2019xny}, and the 
FCC-ee is expected to deliver $10^{12}$ $Z$s
(``TeraZ'')~\cite{Blondel:2019yqr}); by searching for new particles
responsible for the deviations, such as leptoquarks or $Z'$ bosons;
by probing the effects of Wilson coefficients in the kinematical
distributions sensible to virtual effects; and by improving the
precision of the observables that enter our global fits. Due to the
high number of $Z$ bosons produced, EW observables are a prime example
of the advantages of $e^+e^-$ colliders. 

In what follows, we will focus on the prospects of indirect discovery
using Wilson coefficients and EW observables. The increased
center-of-mass energy of the future colliders improves the sensitivity
to the effects of any dimension-6 Wilson coefficient. This is evident 
from the energy scaling of the $2\to 2$ scattering amplitudes,
$A_6 \propto \frac{E^2}{\Lambda^2}\,$~\cite{Maltoni:2019aot}.
\begin{figure}
\centering
\includegraphics[width=0.8\textwidth]{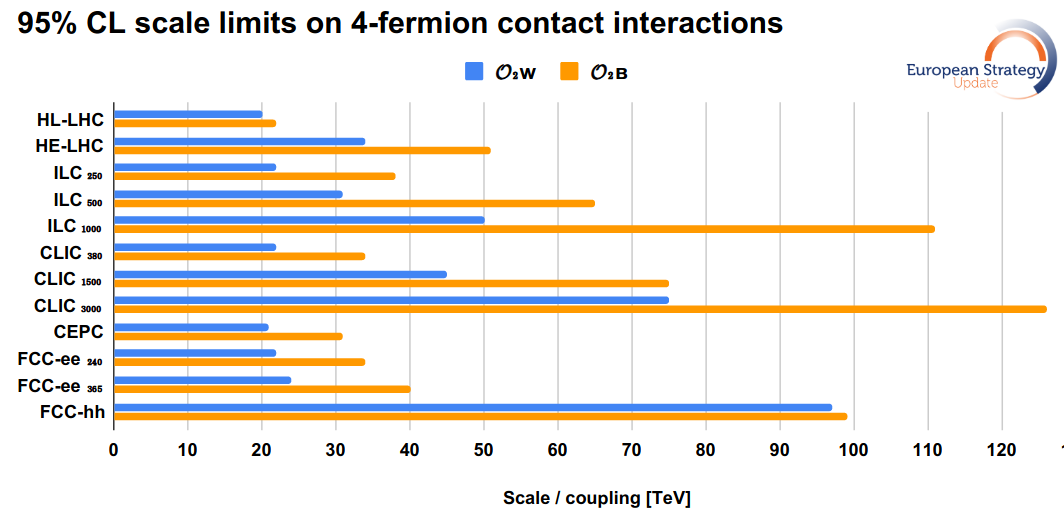}
\caption{95\% exclusion reach in future colliders from the operators
  $\mathcal{O}_{2W}$ (blue) and $\mathcal{O}_{2B}$ (orange). The effective scale
  is given by $\Lambda/({g'}^2 \sqrt{C_{2W}})$ for the blue bars, and
  $\Lambda/(g^2 \sqrt{C_{2B}})$ for the orange bars. Taken from~\cite{Strategy:2019vxc}.}
\label{fig:scales} 
\end{figure}

The study of neutral-current benefits greatly from the clean signatures
and small theoretical uncertainties provided by lepton colliders. The
use of polarized beams allows for the study of the different helicity
structures of the Wilson coefficients. The constraints from lepton colliders
for the four-fermion contact operators are the result of a variety of final
states. For example, the $e^+ e^- \to t \overline{t}$ events can constrain
$C_{\ell q(1)}-C_{\ell q(3)}$, while $e^+ e^- \to b \overline{b}$ events can constrain
$C_{\ell q(1)}+C_{\ell q(3)}$~\cite{Durieux:2018tev}. Also the leading 
higher-derivative corrections to the $W$ and $Z$ bosons propagators from
the $\mathcal{O}_{2W}$ and $\mathcal{O}_{2B}$ operators, 
\begin{equation}
  \mathcal{O}_{2W} = (D^\mu W_{\mu\nu})^i(D_\rho W^{\rho\nu})^i, \qquad\qquad
  \mathcal{O}_{2B} = (\partial^\mu B_{\mu \nu})(\partial_\rho B^{\rho \nu})
\end{equation}
from the Strongly-Interacting Light Higgs (SILH)
basis~\cite{Giudice:2007fh} can be recast into flavour-universal
four-fermion operators using the equations of motion
\begin{equation}
  \mathcal{O}_{2W} = -\frac{g^2}{4} \sum_{i,j} O_{\ell q(3)}^{iijj}+\cdots,\qquad\qquad
  \mathcal{O}_{2B} = -\frac{{g'}^2}{6}\sum_{i,j}O_{\ell q(1)}^{iijj}+\cdots,
\end{equation}
where $g$ and $g'$ are the gauge couplings for the $SU(2)_L$ and $U(1)_Y$ SM groups.

The exclusion reach for the operators $\mathcal{O}_{2W}$ and
$\mathcal{O}_{2B}$ in the different colliders are depicted in
Figure~\ref{fig:scales}, taken from~\cite{Strategy:2019vxc}. Lepton colliders
provide better sensitivity for singlet operators ($\mathcal{O}_{2B}$)
than for triplet operators
($\mathcal{O}_{2W}$), while the sensitivity of hadron colliders is
similar in both cases. In its initial stage at $\sqrt{s} = $250 GeV, ILC
is expected to provide a better sensitivity than the high-luminosity
upgrade of LHC. 

An important feature of our model is that it predicts NP 
couplings to electrons similar in magnitude to the couplings to
muons. This opens the option of observation in an $e^+ e^-$ machine,
specially using $e^+ e^- \to b s$ production, which has a very clean SM background, since
this process is only generated at one loop and CKM-suppressed by
$V_{ts}$~\cite{deBlas:2018mhx}. 

The lepton linear colliders running at their initial stages will
generate a great number of $W$ and $Z$ bosons (about $10^8$ in ILC at
$\sqrt{s} =$ 250 GeV and $10^7$ in CLIC at
$\sqrt{s} =$ 380 GeV~\cite{Strategy:2019vxc}). This will allow to improve the precision of
the EW observables: the mass of the $W$ boson $m_W$, and the
decay asymmetries $A$ and rates $R$ of the $Z$ boson. A dedicated
program running at the $Z$ pole would increase the number of bosons by
an order of magnitude, improving accordingly the precision of the
measurements. Circular $e^+ e^-$ colliders using transversely polarised
beams will achieve even better results.

In our fits in section~\ref{sec:fits} we have shown that the EW 
observables, due to the mixing via Renormalization Group Equations,
offer a set of constraints on NP
complementary to those coming from $B$ decays. A significant improvement
in the precision of EW observables would have consequently a great
impact on our results. In order to asses the impact of the improved precision
on our previous analysis, we have performed a new global
fit~\cite{Alda:2021}~{\footnote{Work in preparation. Preliminary results are presented here.}}. 
For the central
values of the EW observables we have used their predictions in our
previous fits~\cite{Alda:2020okk},
and the uncertainty is taken from the ILC at $\sqrt{s} =$ 250 GeV projections
from~\cite{Strategy:2019vxc}. The assumed values of the central EW
observables and their uncertainties are shown on
Table~\ref{tab:EWinputsILC}. 
The other observables are unchanged since our
previous work~\cite{Alda:2020okk}. The largest tensions between our
inputs and the SM predictions are
found in the observables $A_e$ and $m_W$, being $5.6\, \sigma$ and
$2.9\, \sigma$ respectively.
\begin{table}
\centering
\begin{tabular}{|c|c|c|c|c|c|c|}\hline
\multirow{2}{*}{Obs} & \multicolumn{5}{|c|}{Central value} & \multirow{2}{*}{Error} \\ \cline{2-6}
& IV & V & VI & VII & IX & \\\hline
$m_W$ [GeV] & 80.363 & 80.382 & 80.342 & 80.365 & 80.359 & 0.002 \\\hline
$A_e$ & 0.14779 & 0.14900 & 0.14580 & 0.14785 & 0.14738 & 0.00015 \\\hline
$A_\mu$ & 0.1471 & 0.1488 & 0.1457 & 0.14716 & 0.1467 & 0.0008 \\\hline
$A_\tau$ & 0.1474 & 0.1494 & 0.1463 & 0.14798 & 0.1474 & 0.0008 \\\hline
$A_c$ & 0.6677 & 0.6683 & 0.6670 & 0.6677 & 0.6675 & 0.0014 \\\hline
$A_b$ & 0.9347 & 0.9349 & 0.9346 & 0.9348 & 0.9347 & 0.0006 \\\hline
$R_e$ & 20.73 & 20.73 & 20.73 & 20.73 & 20.73 & 0.02 \\\hline
$R_\mu$ & 20.74 & 20.74 & 20.74 & 20.74 & 20.74 & 0.02 \\\hline
$R_\tau$ & 20.78 & 20.77 & 20.77 & 20.77 & 20.77 & 0.02 \\\hline
$R_c$ & 0.1722 & 0.1722 & 0.1722 & 0.1722 & 0.1722 & 0.0008 \\\hline
$R_b$ & 0.2158 & 0.2158 & 0.2158 & 0.2158 & 0.2158 & 0.0002 \\\hline
\end{tabular}
\caption{Assumed central values for the EW observables and their
  uncertainties used in the ILC global fits for
  several scenarios.}
\label{tab:EWinputsILC}
\end{table}

\begin{figure}
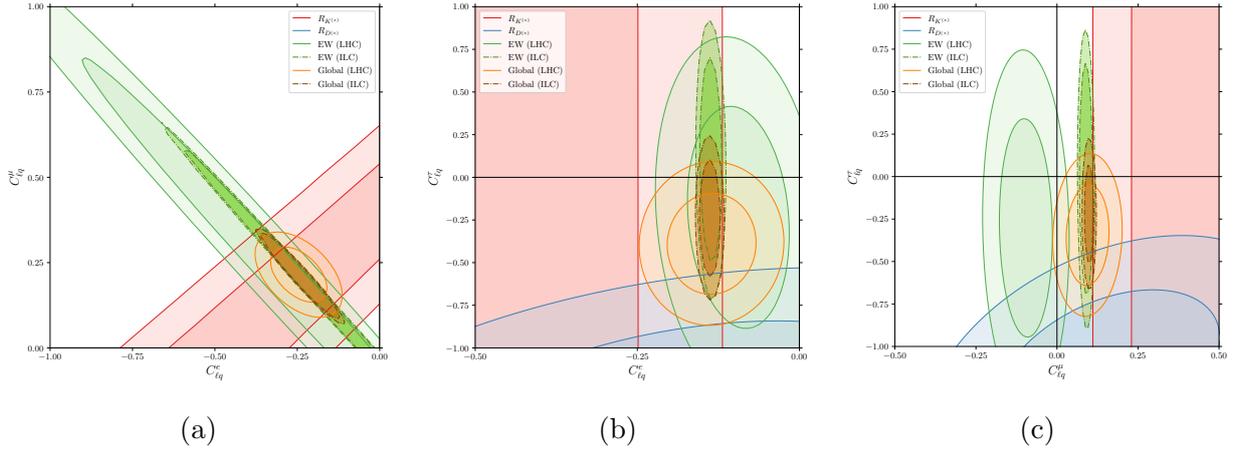

\begin{center}
\begin{tabular}{ccc}
\resizebox{0.3\textwidth}{!}{\input{scIV_EW_zoom.pgf}}&
\resizebox{0.3\textwidth}{!}{\input{scV_EW_zoom.pgf}}&
\resizebox{0.3\textwidth}{!}{\input{scVI_EW_zoom.pgf}}\\
(a)&(b)&(c)
\end{tabular}
\caption{Detail of the $1\sigma$ and $2\sigma$ contours for scenarios with two lepton flavours present 
  in the Wilson coefficients: (a) Scenario IV, (b) Scenario V, and (c) Scenario VI. Solid
  lines correspond to the current fits, and dash-dotted lines to the fits including the ILC
  projections.}
\label{im:fitsILC}
\end{center}
\end{figure}
The fits to scenarios IV, V and VI using the projected ILC values
are already included in Figure~\ref{im:globalfits}. For clarification,
a detailed region in which the ILC prediction appears is 
displayed in Figure~\ref{im:fitsILC}. The LFU-conserving direction of the fit,
corresponding to the linear combination $C_2 \sim (-\Clq^e -
\Clq^\mu)/\sqrt{2}$, is even more tightly
constrained due to the better precision of the EW observables, obtaining
$C_2 = -0.034\pm 0.011$ in scenario VII. The LFUV direction of the fit
remains unchanged, since the
EW observables are not sensitive to these deviations.

\begin{table}
\centering
\begin{tabular}{|c|c|c|c|}\hline
Observable & Scenario IV & Scenario VII & Scenario IX \\\hline
$R_K^{[1.1, 6]}$ & $0.802 \pm 0.003$ & $0.803 \pm 0.005$ & $0.807 \pm 0.004$ \\\hline
$R_{K^*}^{[0.045,\ 1.1]}$ & $0.872 \pm 0.004$ & $0.872 \pm 0.008$ & $0.873 \pm 0.009$ \\\hline
$R_{K^*}^{[1.1,\ 6]}$ & $0.804 \pm 0.005$ & $0.805 \pm 0.006$  & $0.809 \pm 0.008$ \\\hline
$R_D^\ell$ & $0.302 \pm 0.005$ & $0.309 \pm 0.005$ & $0.310 \pm 0.005$ \\\hline
$R_{D^*}^\ell$ & $0.254 \pm 0.003$ & $0.260 \pm 0.003$ & $0.261 \pm 0.004$ \\\hline
$R_{D^*}^\mu$ & $0.262 \pm 0.003$ & $0.267 \pm 0.004$ & $0.269 \pm 0.004$ \\\hline
\end{tabular}
\caption{Values of the $\RKp$ and $\RDp$ observables in the scenarios with better pulls for the fit with the upgraded ILC precision.}
\label{tab:observables_ILC}
\end{table}
\begin{figure}
\begin{center}
\begin{tabular}{cc}
\includegraphics[width=0.553\textwidth]{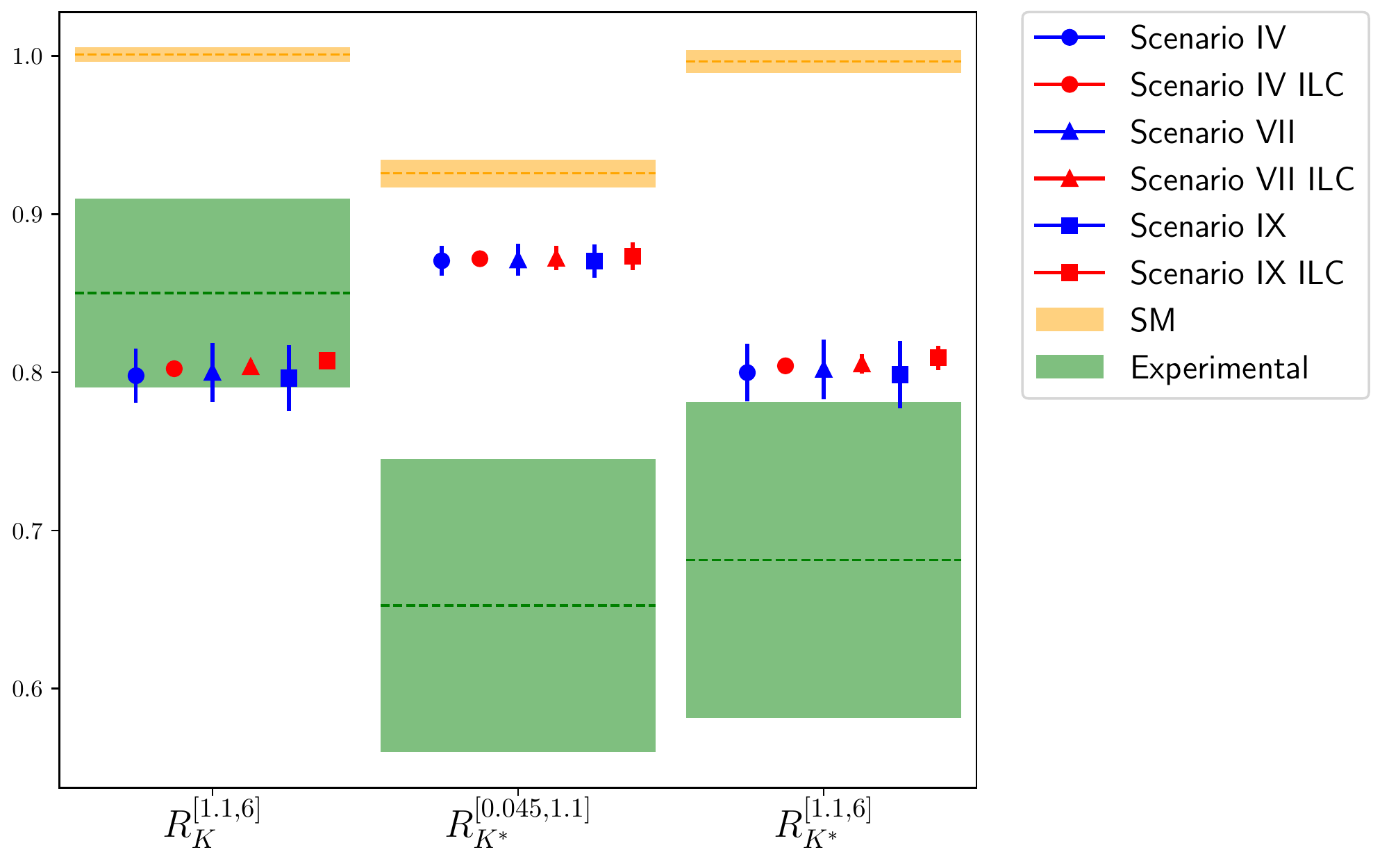}&
\includegraphics[width=0.397\textwidth]{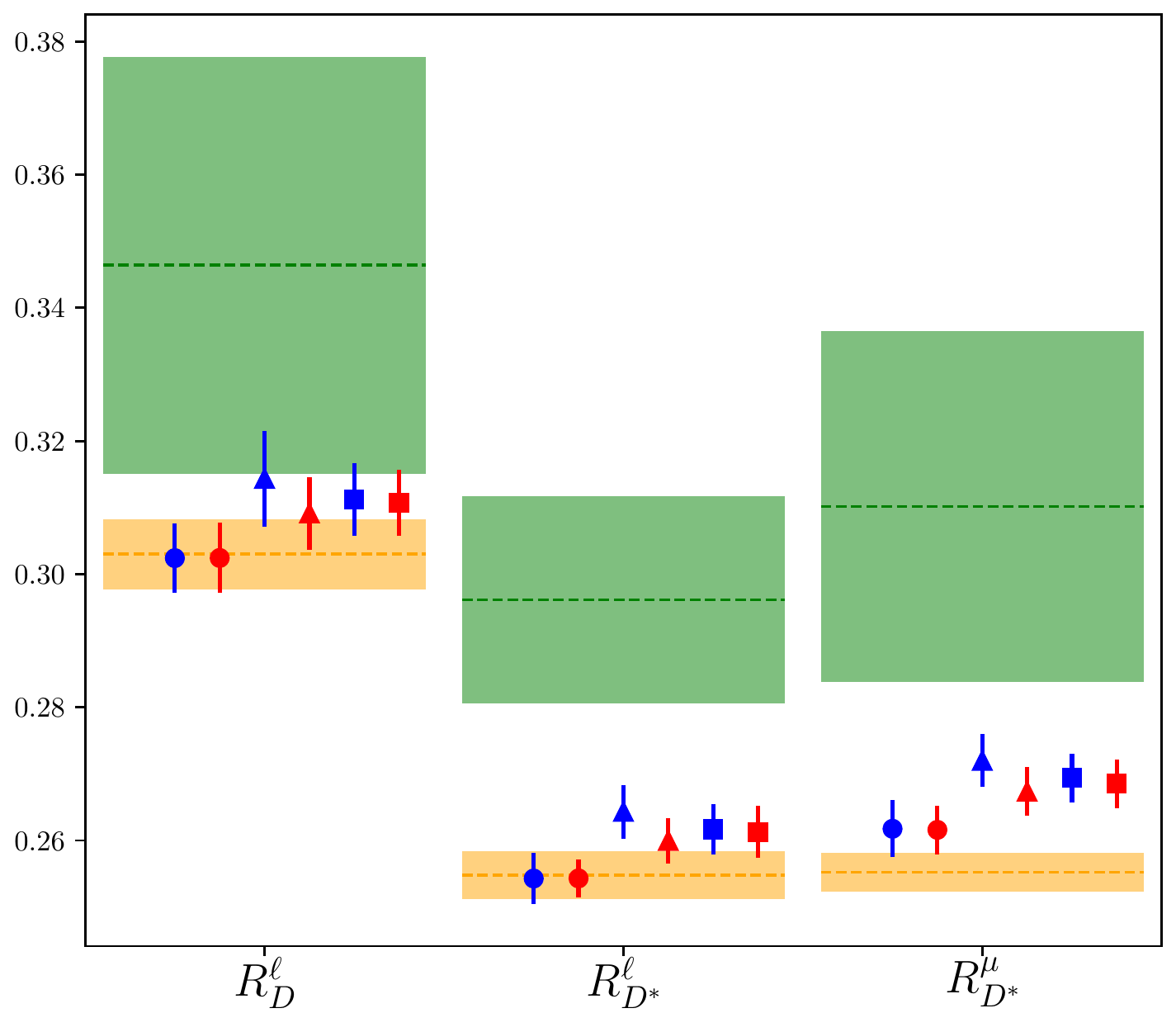}\\
(a)&(b)
\end{tabular}
\caption{Central value and $1 \sigma$ uncertainty of the (a) $\RKp$
  observables, and (b) $\RDp$ observables in scenarios 
IV, VII and IX (blue lines for current predictions, red lines for ILC-based predictions), compared to the SM prediction (yellow) and experimental measurements (green).}
\label{im:RK_ILC}
\end{center}
\end{figure}
The predictions for the \RKp and \RDp observables in the best fit points
for scenarios IV, VII and IX with the upgraded ILC precision can be found in
Table~\ref{tab:observables_ILC}. Clearly, the precision in all those observables is
improved. To compare with our previous fit, Figure~\ref{im:RK_ILC}
displays the central value and
$1 \sigma$ uncertainty of the $\RKp$ and $\RDp$ observables in the above mentioned scenarios
for the current predictions (blue lines) and the ILC predictions (red lines). 
The error in all those observables is now dominated by the theoretical
uncertainty, as a consequence of the reduction of the allowed region for the Wilson
coefficients in the fits. The error of the $\RKp$ observables is improved
up to factor of 3, specially in the $1.1 < q^2 < 6$ region, in which the
results of the global fits are in agreement
with the experimental measurements.

\section{Conclusions}\label{sec:conclu}
Several measurements of $B$ meson decays performed in the recent years
indicate a possible violation of Lepton Universality that may represent an 
indirect signal of New Physics. In this note we summarise the results obtained
in~\cite{Alda:2018mfy,Alda:2020okk} for the
analysis of the effects of the global fits to the Wilson
coefficients assuming a model independent effective Hamiltonian approach.
The global fit includes $b\to s \mu\mu$ observables (including the
Lepton Flavour Universality ratios $\RKp$, the angular observables $P_5'$ and the branching ratio of
$B_s \to \mu\mu$), as well as the $\RDp$, $b \to s \nu \bar{\nu}$ and electroweak precision
observables ($W$ and $Z$ decay widths and branching ratios to leptons).

We consider different scenarios for the
phenomenological analysis such that New Physics is present in one, two or three of the Wilson
coefficients at a time. For all scenarios 
we compare the results of the global fit with respect to both the SM and the more
general scenario: the best fit point of the three
independent Wilson coefficients scenario in which New Physics modifies 
each of the operators independently.

We conclude that, when New Physics contributes to only one
lepton flavour operator at a time, the largest pull from the Standard
Model prediction, almost $3\ \sigma$,
appears when the coupling to electrons
is added independently, corresponding to our scenario I. 
In those scenarios in which New Physics is present in two of the Wilson 
coefficients simultaneously, the best fit corresponds to the case of 
scenario IV, where the contributions to $\Clq^e$ and $\Clq^\mu$ are
favoured with a pull of $4.97\ \sigma$ with respect to the SM.
If we focus on the more general scenario of three independent Wilson
coefficients, we found that the prediction of the $\RDp$ and $\RKp$ observables is
improved in the scenario in which the three $\Clq$ operators receive independent NP
contributions: Scenario VII. In this case, the pull from the Standard Model is
$4.97\ \sigma$ and the predictions for the $\RKp$ observables are very similar to
the case of Scenario IV.  A better fit to $\RDp$ observables,
and specially to $R_D^\ell$, is obtained in this scenario. We
also found that Scenario IX provides a similar fit goodness with a smaller set
of free parameters.

Finally, we have discussed that the future particle colliders, and in
particular the linear lepton colliders ILC and CLIC, will provide
valuable new information to cast light on the $B$ anomalies.
For the $\RKp$ observables, the error is improved up to factor of 3,
specially in the $1.1 < q^2 < 6$ region, in which the
results of the global fits are in agreement
with the experimental measurements. Improved precision in electroweak observables will help
constrain the global fits in a complementary way to $B$-physics
experiments. 

\section*{Acknowledgements}

The work of J.~A. and S.~P. is partially supported by Spanish grants 
MINECO/FEDER grant FPA2015-65745-P, PGC2018-095328-B-I00 
(FEDER/Agencia estatal de investigaci{\'o}n) and DGIID-DGA No. 2015-E24/2.
J.~A. is also supported by the 
Departamento de Innovaci\'on, Investigaci\'on y Universidad of Arag\'on
goverment, Grant No. DIIU-DGA. 
J.G. has been suported by MICIN under projects PID2019-105614GB-C22 and 
CEX2019-000918-M of ICCUB (\textit{Unit of Excellence Mar{\'\i}a de Maeztu 2020-2023})
and AGAUR (2017SGR754).

\end{document}